\documentclass{sf2a-conf}
\usepackage{graphicx}
\usepackage{natbib}
\bibpunct{(}{)}{;}{a}{}{,}
\newcommand{\aap}{A\&A}
\newcommand{\apj}{ApJ}

\newcommand{\tableline}{\hline}

\begin{document}

\TitreGlobal{SF2A 2008}

\title{The luminosity of GRB afterglows as distance estimator}
\author{Gendre, B.}\address{Laboratoire d'Astrophysique de Marseille/CNRS/Université de Provence}
\author{Galli, A.}\address{IASF-Roma}
\author{Bo\"er, M.}\address{Observatoire de Haute-Provence/CNRS}
\runningtitle{GRB afterglows distance estimator}
\setcounter{page}{237}

\index{Gendre, B.}
\index{Galli, A.}
\index{Bo\"er, M.}

\maketitle
\begin{abstract}We investigate the clustering of afterglow light curves observed at X-ray and optical wavelengths. We have constructed a sample of 61 bursts with known distance and X-ray afterglow. GRB sources can be divided in three classes, namely optical and X-ray bright afterglows, optical and X-ray dim ones, and optically bright -X-ray dim ones. We argue that this clustering is related to the fireball total energy, the external medium density, the fraction of fireball energy going in relativistic electrons and magnetic fields. We propose a method for the estimation of the GRB source redshift based on the observed X-ray flux one day after the burst and optical properties. We tested this method on three recently detected SWIFT GRBs with known redshift, and found it in good agreement with the reported distance from optical spectroscopy.  \end{abstract}
%
\section{Introduction}

Long Gamma-Ray Bursts (GRBs) \citep[for a review, see][]{mes06} are linked to the final stages of the stellar evolution. As these objects are extremely luminous and can be detected up to large distance, they are interesting for studies of cosmology in the redshift range 1-15. However, the use of GRB for cosmological studies needs to build a robust indicator of their distance, whenever possible based on their intrinsic properties.

Hints of standardization of the X-ray afterglow luminosities were first discovered by \citet{boe00}, who found evidences for clustering in the X-ray luminosity of BeppoSAX afterglows, and confirmed later by \citet{gen05}. In the following we will refer to these articles respectively as paper I and II. This study was completed by \citet{nar06} and \citet{lia06} who found independently that optical afterglows were also clustered in luminosity, and by \citet{gen08b} who extended this study towards infrared wavelengths. We have tried to derive a method for estimating the burst distance based on that property. In Section \ref{sec_ana} we present the luminosity clusterings. In section \ref{sec_redshift} we present our method of GRB distance estimation from the X-ray afterglow light curve. We test this method on several GRBs detected recently by SWIFT, and we propose an estimation of the redshift for two previously detected GRBs of unknown distance.

\section{Luminosity clustering}
\label{sec_ana}

\subsection{X-ray clustering}
\begin{figure*}
   \centering
   \includegraphics[width=8.4cm]{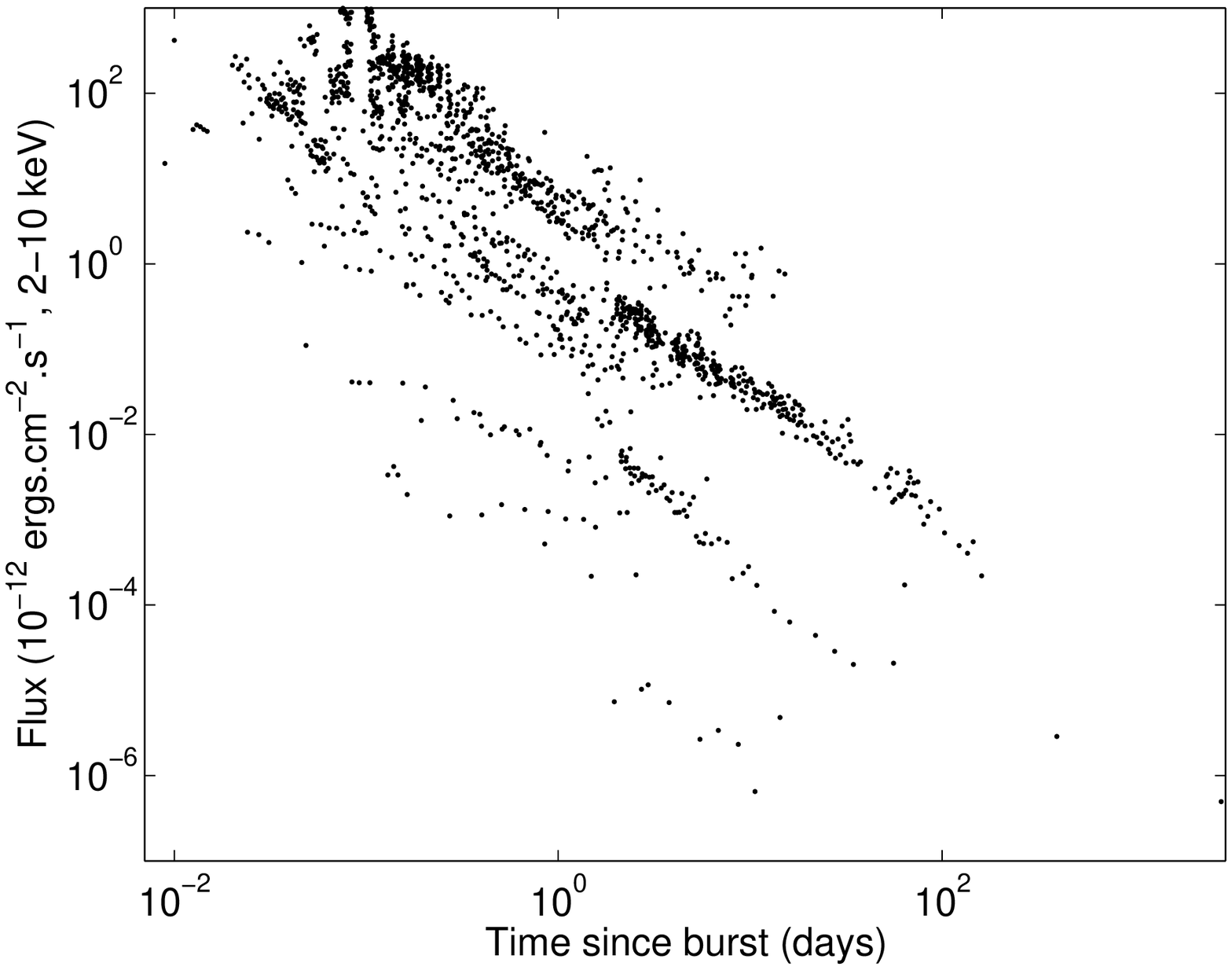}
   \includegraphics[width=8cm]{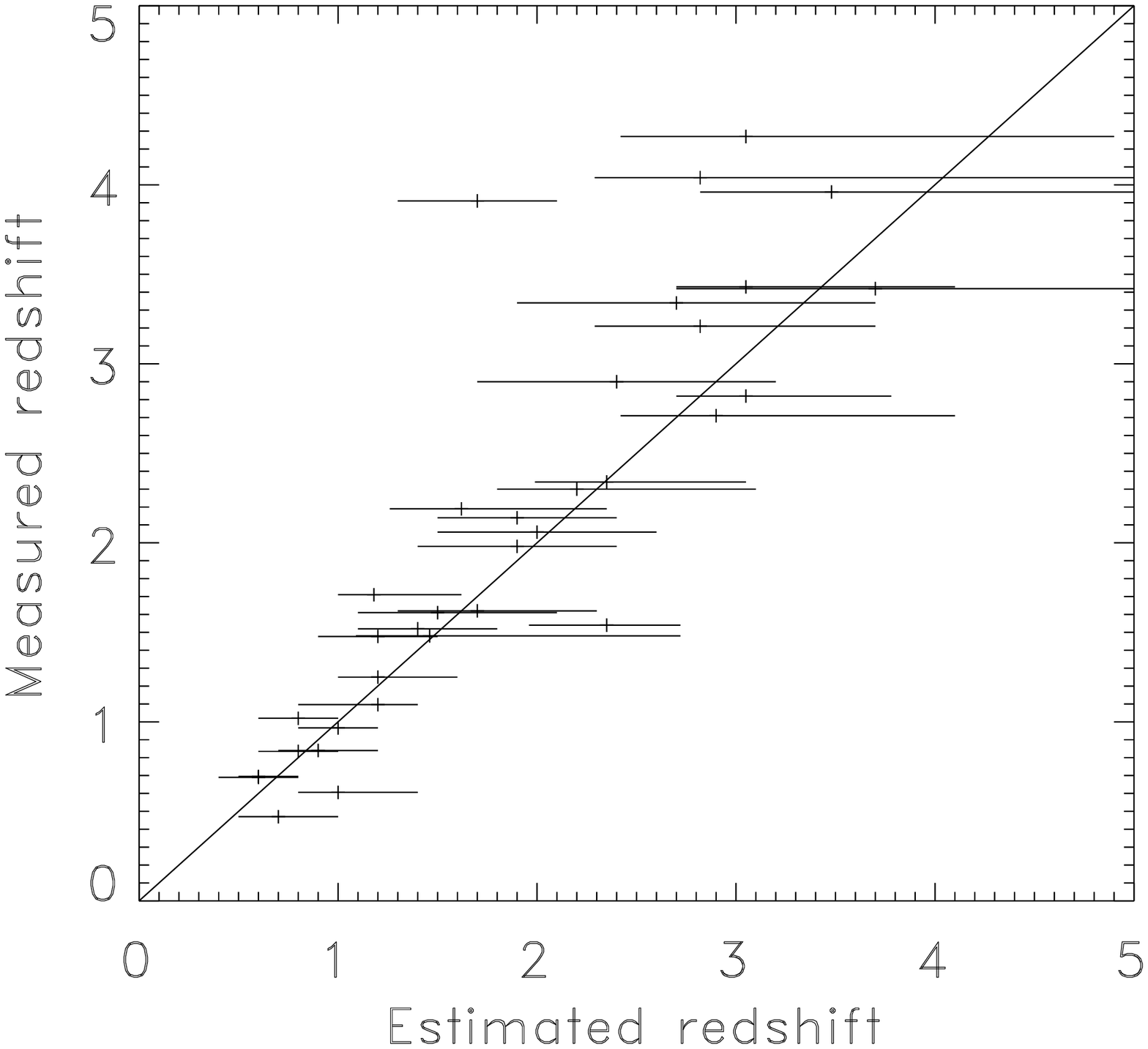}
      \caption{Left: the light curves of our sample, corrected for distance effects (see text). The groups reported in papers I and II are clearly seen. Right: Comparison of the estimated redshift versus the measured one for the bursts defining our sample. The solid line corresponds to equality.
              }
         \label{fig2}
   \end{figure*}

Our sample of GRBs with known redshift and X-ray afterglow observations includes all afterglows observed by BeppoSAX, XMM-Newton, Chandra, and SWIFT prior to the 1st of August 2006 with a $t_{90}$ larger than 2.0 seconds (in order to exclude short bursts). Data analysis is explained in detail in \citet{gen08}. The results are displayed in Fig. \ref{fig2}. The groups reported in papers I and II are apparent. We note however some dispersion during the first part of the light curves. We interpret this as a consequence on the error on the $T_a$ measurement. We refer to the bright group as {\it xI} and the dimmer one as {\it xII}, as in paper II. Height bursts do not follow this relation. We consider that these low luminosity events form a specific group, referred as {\it xIII} in the following.

\subsection{The optical afterglows}

\citet{nar06} and \citet{lia06} have shown that optical afterglow light curves also display a clustering effect. They found two groups, and we will refer to them as {\it oI} and {\it oII} for the bright and dim group respectively. While an {\it oI} burst can belong to {\it xI} or {\it xII}, {\it oII} bursts are dim both at optical and X-ray wavelengths. 

\subsubsection{The nature of the clustering}

The optical and X-ray data define 3 groups : {\it xI-oI} events, {\it xII-oI} events, {\it xII-oII} events, plus the outliers {\it xIII} events. We propose that the presence of different groups is ascribed to different families of $\epsilon_e$ and $\epsilon_B$ values:
\begin{itemize}
\item a family of 'magnetized fireball' that produces the group {\it xII-oI}. In such a case the fireball transfers only a low fraction of its energy into relativistic electrons. 
\item a family of 'less magnetized fireball' that produces the groups {\it xI-oI} and {\it xII-oII}. The fraction of total energy going into magnetic fields is roughly one order of magnitude lower than group {\it xII-oI}. These two groups can be related to an high and low fraction of energy going in relativistic electrons, respectively. 
\end{itemize}

\section{Estimation of GRB source distance}
\label{sec_redshift}

To date, most of the redshift measurements made on GRB afterglows were done by optical spectroscopic or photometric observations. However, not all GRBs can be followed in optical. On the other hand, after the launch of the SWIFT satellite, nearly all GRBs have an homogeneous X-ray follow-up. Hence, a distance measurement method based solely on X-ray observations could be very interesting if one wants to use a large sample of GRB sources for cosmological studies. We used the observed luminosity clustering in X-ray to derive a distance estimator. Table \ref{table_estimate}, which provides the redshift needed to comply with the relation for both groups for a given observed flux, is directly usable to estimate that estimator. Note that the belonging group can be fixed only through broad-band modeling. In practical, this leads to two redshift estimates. If the bursts has an optical afterglow there is a possible way to decide which of the estimate is valid: if a GRB afterglow is observed by UVOT (or in the B band for ground based telescopes) then it cannot belong to group I, and the ambiguity on the redshift determination is cleared. 

We calibrated this method by deriving the estimated redshift for the bursts of our sample (for which the group is known), and comparing this value with the measured redshift. The results are displayed in Fig. \ref{fig2}. The estimated redshift agrees with the measured one for most of the bursts. The only discrepancies arises at low redshift: as a conservative approach we prefer to restrict the validity of our method to source located at redshifts larger than 0.5.

\begin{table}

\caption{Flux to redshift conversion. This table should be used as an estimate of the redshift for bursts observed by any X-ray observatory. The flux is given in the 2-10 keV band at 1 day after the burst (observer frame). The redshift has been calculated for an energy index of 1.2, the uncertainty is 30\%.\label{table_estimate}}
\centering                          
\begin{tabular}{c c c }        
\tableline\tableline                 
Flux & Group I redshift & group II redshift \\    
(erg s$^{-1}$ cm$^{-2}$)&&                  \\
\tableline                        
$1 \times 10^{-14}$ & ---    &  4.43  \\
$2 \times 10^{-14}$ & ---    &  3.28  \\
$3 \times 10^{-14}$ & ---    &  2.72  \\
$4 \times 10^{-14}$ & ---    &  2.35  \\
$5 \times 10^{-14}$ & ---    &  2.12  \\
$6 \times 10^{-14}$ & 7.80   &  1.96  \\
$7 \times 10^{-14}$ & 7.09   &  1.83  \\
$8 \times 10^{-14}$ & 6.38   &  1.75  \\
$9 \times 10^{-14}$ & 6.05   &  1.68  \\
$1 \times 10^{-13}$ & 5.78   &  1.62  \\
$2 \times 10^{-13}$ & 4.10   &  1.26  \\
$3 \times 10^{-13}$ & 3.48   &  1.09  \\
$4 \times 10^{-13}$ & 3.05   &  0.96  \\
$5 \times 10^{-13}$ & 2.82   &  0.89  \\
$6 \times 10^{-13}$ & 2.59   &  0.82  \\
$7 \times 10^{-13}$ & 2.42   &  0.77  \\
$8 \times 10^{-13}$ & 2.29   &  0.73  \\
$9 \times 10^{-13}$ & 2.10   &  0.69  \\
$1 \times 10^{-12}$ & 1.99   &  0.66  \\
$2 \times 10^{-12}$ & 1.50   &  0.50  \\
$3 \times 10^{-12}$ & 1.29   &  ---  \\
$4 \times 10^{-12}$ & 1.16   &  ---  \\
$5 \times 10^{-12}$ & 1.06   &  ---  \\
$6 \times 10^{-12}$ & 0.99   &  ---  \\
$7 \times 10^{-12}$ & 0.94   &  ---  \\
$8 \times 10^{-12}$ & 0.89   & ---   \\
$9 \times 10^{-12}$ & 0.83   & ---   \\
$1 \times 10^{-11}$ & 0.79   &  ---  \\
$2 \times 10^{-11}$ & 0.56   &  ---  \\
$3 \times 10^{-11}$ & 0.50   &  ---  \\

\tableline                                   
\end{tabular}
\end{table}

\begin{table}
\caption{Redshift estimates derived from the relation using either the group I and II hypotheses for pre-SWIFT bursts without known distance.\label{table2}}
\centering                          
\begin{tabular}{ccc}
\tableline
\tableline
GRB  name  &  Redshift           & estimate     \\
           &  group I            & group II     \\
\tableline
GRB 980329 & 4.2 $\pm$ 1.2       & 1.2 $\pm$ 0.2\\
GRB 980519 & 3.8 $\pm$ 0.7       & 1.4 $\pm$ 0.2\\
GRB 990704 & 3.5 $\pm$ 0.9       & 1.3 $\pm$ 0.3\\
GRB 990806 & 4.7$^{+1.6}_{-0.7}$ & 1.6 $\pm$ 0.3\\
GRB 001109 & 2.3 $\pm$ 0.7       & 0.8 $\pm$ 0.2\\
GRB 001025A& 5.8 $\pm$ 1.8       & 2.2 $\pm$ 0.4\\
GRB 020322 & 5.0 $\pm$ 1.5       & 1.5 $\pm$ 0.3\\
GRB 040106 & 3.4 $\pm$ 0.5       & 1.0 $\pm$ 0.2\\
GRB 040223 & 5.5$^{+2.0}_{1.2}$  & 1.7 $\pm$ 0.2\\
GRB 040827 & 8.0 $\pm$ 2.0       & 1.9 $\pm$ 0.3\\
\tableline
\end{tabular}
\end{table}

The distance estimation of several events with no known redshift, good temporal sampling and spectral informations is presented in table \ref{table2}. GRB 980519 was observed in U band \cite{jau01}. As the Lymann $\alpha$ cut-off cross the U band at $z \sim 2.8$, this burst should not have been observed in the high distance hypothesis. U observations can indeed solve the problem of group classification, and we propose a redshift measurement for GRB 980519 of 1.4 $\pm$ 0.2. For the same reason, GRB 040827 has $z = 1.9 \pm 0.3$, since an optical afterglow has been observed \citep{mel04}.

\begin{table}
\caption{Redshift estimates derived from the relation using either the group I and II hypotheses for SWIFT bursts with known distance.\label{table_sup}}
\centering                          
\begin{tabular}{cccc}
\tableline
\tableline
GRB  name           &  Group I  & Group II & Measured   \\
                    &  estimate & estimate & redshift  \\
\tableline
GRB 070529 &  7.8      &  1.96    & 2.44 \\
GRB 070611 &  6.05     &  1.68    & 2.04 \\
GRB 070721B&  ---      &  3.28    & 3.62 \\
\tableline
\end{tabular}
\end{table}

Several GRB were recently detected by SWIFT and their distance derived using optical spectroscopy. We thus have a sample of bursts which allows to test this correlation independently. To simulate actual conditions for an unknown event we used the SWIFT count rate light curve available from their web site, and we apply a mean spectral index of 1.2 for the count-to-flux conversion. The results are indicated in Table \ref{table_sup}, indicating an agreement in accordance with the error estimated previously.

\section{Conclusions}

We have investigated the clustering of afterglow light curves observed previously in X-ray and in optical. Adding SWIFT bursts to the previous sample reported in paper II, we still confirm our previous findings. On a sample of 61 events the X-ray light curves cluster in two groups, with a significance larger than 6 $\sigma$. We compared the classification within each group in X-ray and in optical, and found three classes: {\it bright} optical and X-ray afterglows, {\it dim} ones, and optically {\it bright}-X-ray {\it dim} ones. 

Using the observed X-ray afterglow properties of GRBs, we propose a new, simple, method for the determination of the source distance based on X-ray data; optical photometry in U and V bands may help to clear the degeneracy between the two estimates found. We apply this method to a sample of GRB source of unknown redshift. We propose an estimation of the redshift for GRB 980519 (1.4 $\pm$ 0.2) and for GRB 040827 ($1.9 \pm 0.3$).

\acknowledgements
BG acknowledge support from COFIN grant 2005025417, and from the Centre Nationnal d'Etudes Spatiales. We are pleased to thank Alessandra Corsi, Giulia Stratta. This work made use of data supplied by the UK Swift Science Data Centre at the University of Leicester.

\end{document}